\documentstyle[12pt, fleqn]{article}

\begin{document}

\author{{\bf Shahin S. Agaev}\\
{\it High Energy Physics Lab., Baku State University,}\\
{\it Z.Khalilov st. 23, 370148 Baku, Azerbaijan}}
\title{{{\bf LIGHT MESONS ELM FORM FACTOR AND RUNNING COUPLING EFFECTS}}\thanks
{Talk given at the Euroconference QCD 98, Montpellier 2-8th July 1998, France, 
to appear in Proceedings.}} 
\date{}
\maketitle
\begin{abstract}
The pion and kaon electromagnetic form factors $F_{M}(Q^2)$ are calculated
at the leading order of pQCD using the running coupling constant method.
In computations dependence of the mesons distribution amplitudes on the hard scale
$Q^2$ is taken into account. The Borel transform and resummed expression for
$F_M(Q^2)$ are found. The effect of the next-to-leading order term in 
expansion of $\alpha_S(\lambda Q^2)$ in terms of $\alpha_S(Q^2)$ on the pion form
factor $F_{\pi}(Q^2)$ is discussed, comparison is made with the infrared 
matching scheme's result.
\end{abstract}
\newpage

{\bf 1.} Investigation of the infrared (ir) renormalon effects in various 
inclusive and exclusive processes is one of the important and interesting 
problems in the pQCD~\cite {br} (and references therein). It is known that 
all-order resummation of ir renormalons corresponds to
the calculation of the one-loop Feynman diagrams with the running coupling constant
$\alpha_{S}(-k^{2})$ at the vertices or, alternatively, to calculation of the
same diagrams with non-zero gluon mass. Both these approaches are generalization
of the Brodsky, Lepage and Mackenzie (BLM) scale-setting method \cite {blm} and amount
to absorbing certain vacuum polarization corrections appearing at higher-order
calculations into the one-loop QCD coupling constant. Studies of ir renormalon
problems have opened also new prospects for evaluation of higher twist corrections 
to processes' characteristics.

Unlike inclusive processes exclusive ones have additional source of ir renormalon
contributions \cite {ag1}-\cite {ag3}; integration over longitudinal fractional momenta of
hadron constituents in the expression of the elm form factor generates such
corrections.

{\bf 2.} In the context of pQCD the meson elm form factor has the form,

\begin{equation}
F_M(Q^2)=\int_0^1 \int_0^1 dxdy\phi _M^{*}(y,\hat{Q}^2)
T_H(x,y;Q^2,\alpha _S(\hat{Q}^2))
\phi _M(x,\hat{Q}^2),\label{1} 
\end{equation}
where $Q^2=-q^2$ is the square of the virtual photon's four-momentum. 
Here $\phi_M$ is the meson distribution amplitude, 
containing all non-perturbative hadronic binding effects, whereas 
$T_H$ is the hard-scattering amplitude 
of the subprocess $q \overline{q}'+\gamma^{*}
\rightarrow q\overline{q}'$ and can be found using pQCD. 
In (1) $\hat{Q}^2$ is the factorization and renormalization scale, which is 
taken as the square of the momentum transfer of the exchanged hard gluon in 
corresponding Feynman diagrams. Such choice for $\hat{Q}^2$ allows one to 
remove large terms proportional to $\ln(Q^2/\mu_{F}^2)$ and $\ln(Q^2/\mu_{R}^2)$ 
from the next-to-leading order correction to $T_{H}$ \cite {braat}.

At the leading order
$T_H$ is given by the following expression 

\begin{equation}
T_H=\frac{16\pi C_F}{Q^2}\left[ \frac 23 \frac{\alpha _S(Q^2(1-x)(1-y))}{(1-x)(1-y)}
+\frac 13\frac{\alpha _S(Q^2xy)}{xy}\right],~~C_{F}=\frac43. \label{2}
\end{equation}

One of the important ingredients of our study is the choice of the meson
distribution amplitude $\phi _M(x,Q^2)$. In the framework of pQCD it is
possible to predict the dependence of $\phi _M(x,Q^2)$ on $Q^2$ using
evolution equation, but not its shape (its dependence on $x$). 
In this work for the pion and kaon we use 
model distribution amplitudes proposed
in Refs.\cite {cz}-\cite {braun}. For the pion they have the following form

\begin{equation}
\phi _\pi (x,\mu _0^2) =\phi _{asy}^\pi (x)\left[a+b(2x-1)^2
+c(2x-1)^4\right] ,\label{3} 
\end{equation}
where $\phi _{asy}^\pi (x)$ is the pion asymptotic distribution amplitude,%
\begin{equation}
\phi _{asy}^\pi (x)=\sqrt{3}f_\pi x(1-x), \label{4}
\end{equation}
and $f_\pi =0.093$~GeV is the pion decay constant.

The constants $a,b,c$ in (3) were found by means of the QCD sum rules
method at the normalization point $\mu _0=0.5$~GeV and take different values
\cite{cz},\cite {fh},\cite {braun} defining the pion's alternative distribution
amplitudes.
For the kaon we have

\begin{equation}
\phi_K(x,\mu_0^2)=\phi_{asy}^{K}(x)\left[a+b(2x-1)^2 
 +c(2x-1)^3 \right], \label{5}
\end{equation}
with $a=0.4,b=3,c=1.25$ and $f_K=0.122$~GeV.

The dependence of the meson distribution amplitude on $Q^2$ can be
obtained by means of the following expression

\begin{equation}
\phi _M(x,Q^2)=\phi _{asy}^{M}(x)\sum^{\infty}_{n=0} 
r_nC_n^{3/2}(2x-1)A_n. \label{6}
\end{equation}
Here $\left\{ C_n^{3/2}(2x-1)\right\} $ are the Gegenbauer polynomials, 
$\gamma _n$ is the anomalous dimension and
$$
A_n=\left [\frac {\alpha_S(Q^2)}{\alpha_S(\mu_{0}^2)} \right]^{\gamma_n/\beta_0}
$$
$\beta_0=11-2n_f/3$ being the QCD beta-function's one-loop coefficient.

The QCD running coupling constant $\alpha_{S}(\hat{Q}^2)$ in Eq.(2) suffers
from ir singularities associated with the behaviour of the 
$\alpha_{S}(\hat{Q}^2)$ in the soft regions $x\to 0,~y \to 0;~~x \to 1,~y \to 1$.
Therefore, $F_{M}(Q^2)$ can be found after proper regularization of
$\alpha_{S}(\hat{Q}^2)$ in these soft end-point regions. 
In the framework of the frozen coupling approximation such regularization is
achieved by equating $\hat{Q}^2$ to its mean value $Q^2/4$ and removing
$\alpha_S(Q^2/4)$ as the constant factor from integrand in (1). As a result,
one obtains form factors with the same shape, but different magnitudes 
depending on a distribution amplitude used in computations. To solve this 
problem in the context of the running coupling method let us relate
the running coupling $\alpha_{S}({\lambda}Q^2)$ in terms of
$\alpha_{S}(Q^2)$ by means of the renormalization group equation \cite{sterm}

\begin{equation}
\alpha _S(\lambda Q^2)\simeq \frac{\alpha _S}{1+\left( \alpha
_S\beta _0/4\pi \right) \ln \lambda } 
 -\frac {\alpha_{S}^2}{4\pi} \frac{\beta_1\ln [1+\left(\alpha_S\beta_0/4\pi\right)\ln \lambda]}
{\beta_0[1+\left(\alpha_S\beta_0/4\pi\right)\ln \lambda]^2}, \label{10}
\end{equation}
where $\alpha_{S}$ is the one-loop QCD coupling constant $\alpha_S(Q^2)$ and
$\beta_1=102-38n_f/3$ is the beta-function's two-loop coefficient.

{\bf 3.} As was demonstrated in our works \cite{ag1}-\cite{ag3}, integration in (1)
using (2) and (7) generates ir divergences and as a result for $F_{M}(Q^2)$
we get a perturbative series with factorially growing coefficients. This
series can be resummed using the Borel transformation \cite {thoft}, 

\begin{equation}
\left[ Q^2F_M(Q^2)\right] ^{res}=\frac{(16\pi f_M)^2}{\beta _0}
\int_{0}^{\infty} du\exp
\left( -\frac{4\pi u}{\beta _0\alpha _S}\right) B\left[ Q^2F_M\right]
(u), \label{11}
\end{equation}
where $B[Q^2F_{M}](u)$ is the Borel transform of the corresponding perturbative 
series \cite {ag2},~\cite {ag3}.

When we take both of the variables $x,y$ in Eq.(2) as the running ones, for
$B[Q^2F_M](u)$ we find

\begin{equation}
B\left[ Q^2F_M\right] (u)=\sum^N_{n=1}\left( \frac{%
{\bf m}_n}{(n-u)^2}+\frac{{\bf l}_n}{n-u}\right) . \label{12}
\end{equation}
The exact expressions for ${\bf m}_n, {\bf l}_n$ can be found in Ref.\cite{ag4}.

The Borel transform (9) has double and single poles at $u=n$. They are ir 
renormalon poles, which are responsible for divergence of the perturbative series
for $Q^2F_M(Q^2)$. The resummed
expression (8) can be calculated with the help of the principal value prescription 
\cite{sterm},\cite{thoft}

\begin{equation}
\left[ Q^2F_M(Q^2)\right] ^{res}=\frac{(16\pi f_M)^2}{\beta _0}\sum^{N}_{n=1}
\left[ -\frac{{\bf m}_n}n 
+({\bf l}_n+{\bf m}_n\ln
\lambda )\frac{li(\lambda ^n)}{\lambda ^n}\right] , \label{13}
\end{equation}
where $li(\lambda)$ is the logarithmic integral \cite{erd},
\begin{equation}
li(\lambda )=P.V.\int_{0}^{\lambda} \frac{dx}{\ln x},~~\lambda =Q^2/\Lambda ^2. \label{14}
\end{equation}

In (10) we have taken into account the dependence of the distribution 
amplitude $\phi_M(x,Q^2)$ on the scale $Q^2$ (in [12], for simplicity, we have
replaced $\phi_M(x,\hat{Q}^2) \rightarrow \phi_M(x,Q^2)$) and it is the generalization of
our results for the pion and kaon elm form factors [4],[5].

In the case with one frozen (for example, $y$) and one running ($x$) variables,
which corresponds to the choice $\hat{Q}^2=Q^2(1-x)/2$ and $\hat{Q}^2=Q^2x/2$ in (2), 
for the pion we find (asymptotic distribution amplitude (4))

\begin{equation}
[Q^2F_{\pi}(Q^2)]^{res}=\frac{(16 \pi f_{\pi})^2}{2\beta_0}\left[\frac
{li(\tilde{\lambda})}{\tilde{\lambda}}-\frac{li(\tilde{\lambda}^2)}{\tilde{\lambda}^2}
\right],
~~\tilde{\lambda}=Q^2/2\Lambda^2. \label{14}
\end{equation}
The ir renormalon analysis carried out (Eqs.(10),(12)) allows one to estimate
power corrections to the light mesons' form factor. Another way of such 
estimation is the ir matching scheme, in the context of which, one 
explicitly divides power corrections from a full expression by introducing moments of 
$\alpha_S$ at low scales as new non-perturbative parameters \cite {web}. 
As an example, let us consider the pion form factor 
($\phi_{asy}^{\pi}(x)$). By freezing one of variables ($y$) we can express
$Q^2F_{\pi}$ in terms of moment integrals $F_p$ defined by the formula
$$
F_p(Q)=\frac{p}{Q^p}\int_0^Qdkk^{p-1} \alpha_S(k^2).
$$
After simple calculations we get
$$
Q^2F_{\pi}(Q^2)=64 \pi f_{\pi}^2 \left \{  \left (\frac{\mu}{Q}\right)^2
F_2(\mu) 
- \left (\frac{\mu}{Q} \right)^4F_4(\mu) \right.
$$ 
\begin{equation}
\left.+\frac{\alpha_S}{4} \left [1- 2 \Gamma(1,2z)+ 
\Gamma(1,4z)\right ] 
+ \frac{\alpha_{S}^{2}\beta_0}{32\pi}\left [3-4\Gamma(2,2z)+
\Gamma(2,4z) \right ] \right \}, \label {15}
\end{equation} 
where $\mu=2$ GeV is the ir matching scale, $z=\ln(Q/\sqrt{2}\mu)$, 
$\alpha_S(Q^2/2)$ and
$\Gamma(n,x)$ is the incomplete gamma function \cite {erd}.

{\bf 4.} Results of our numerical calculations are shown in Fig.1. For 
computation of $Q^2F_{\pi}$ in the context of the ir matching scheme, we
use $N=4$ perturbative estimate for moment integrals $F_p(Q)$ (in Eq.(13),
expression for $N=1$ is written down).

It is worth noting that non-perturbative parameters $F_2(\mu)$ and $F_4(\mu)$
can be found using the running coupling method

$$
F_2(\mu)=\frac{4\pi}{\beta_0}\frac{li(\lambda)}{\lambda},
F_4(\mu)=\frac{8\pi}{\beta_0}\frac{li(\lambda^2)}{\lambda^2},
~~\lambda=\mu^2/\Lambda^2.
$$

For $\Lambda=0.2$ GeV we get $F_2(\mu=2 GeV)=0.421,~F_4(\mu=2 GeV)=0.347$.
In our calculations we take for $F_2(\mu=2 GeV) \simeq 0.5$ the value deduced from 
experimental data and for $F_4$-running coupling prediction $0.347$. As is seen, Eq.(12) and ir matching
scheme give approximately the same results, excluding a region of small $Q^2$.
 
In Fig.2, as an example, the dependence of the ratio $R_K=
[Q^2F_K(Q^2)]^{res}/$ \\ $[Q^2F_K(Q^2)]^0$ on $Q^2$, where $[Q^2F_K(Q^2)]^0$
is the kaon form factor in the frozen coupling approximation, 
is shown. Here we take into account a dependence of the kaon's distribution
amplitude on $Q^2$ (curve {\bf 1}, the dashed curve is taken from Ref.[5],
where this dependence was neglected).
The ir renormalon corrections can be transferred into the scale of 
$\alpha_{S}(Q^2)$ in $[Q^2F_K(Q^2)]^0$ 
\begin{equation}
Q^2\rightarrow e^{f(Q^2)}Q^2,
~~f(Q^2)=c_1+c_2\alpha _S(Q^2)+c_3\alpha_{S}^2(Q^2). \label{15}
\end{equation}
Numerical fitting allows us to get (for $n_f=3$) 
$c_1 \simeq -1.304,c_2 \simeq -35.604, c_3 \simeq 127.25$ 
(curve {\bf 2}). 
The similar results can be obtained also for the pion.

It is interesting to clarify an importance of the second term in (7)
for considering problem. Investigations
demonstrate that an effect of the second term on a whole result is small
(see, Ref.\cite{ag5} for details).
 
{\bf 5.} It is evident that ir renormalon effects enhance the ordinary 
(frozen coupling) perturbative predictions for the pion, kaon elm form factors
approximately two times. Our recent studies confirm that neither a dependence
of $\phi_M(x,Q^2)$ on $Q^2$ nor the next-to-leading order term in Eq.(7) 
changes this picture considerably. This feature of ir renormalons may help
one in solution of a contradiction between theoretical interpretations of 
experimental data for the photon-to-pion transition form factor $F_{\gamma \pi}$
from one side and for the pion elm form factor $F_{\pi}$ from another side. 
Thus in Ref.\cite{av} the authors noted that the scaling and 
normalization of $F_{\gamma \pi}$ to favor of
the pion asymptotic-like distribution amplitude. But then prediction for $F_{\pi}$
obtained using the same amplitude is lower than the data by approximately
a factor of 2. We think that in this discussion a crucial point is a chosen
method of integration in (1). Indeed, unlike $F_{\pi}$ the expression for
$F_{\gamma \pi}$ at the leading order of pQCD does not contain an integration
over $\alpha_S(xQ^2)$. In other words, the running coupling method being
applied to Eq.(1) and to the expression for $F_{\gamma \pi}$ (see, Ref.\cite{av}) 
enhances the perturbative result for the pion elm form factor and
at the same time, does not change $F_{\gamma \pi}$.
This allows us to suppose that in the 
context of pQCD the same pion distribution amplitude may explain experimental data for
both $F_{\gamma \pi}$ and $F_{\pi}$.
\newpage

{\bf FIGURE CAPTIONS}

{\bf Fig.1} The pion elm form factor calculated using $\phi_{asy}^{\pi}(x)$ 
(4). The curves correspond to the following computational schemes: {\bf 1}-
resummed expression (10), {\bf 2}-resummed expression (12), dashed-ir matching
scheme, dot-dashed-frozen coupling approximation.

{\bf Fig.2} The ratio $R_K$ as a function of $Q^2$.

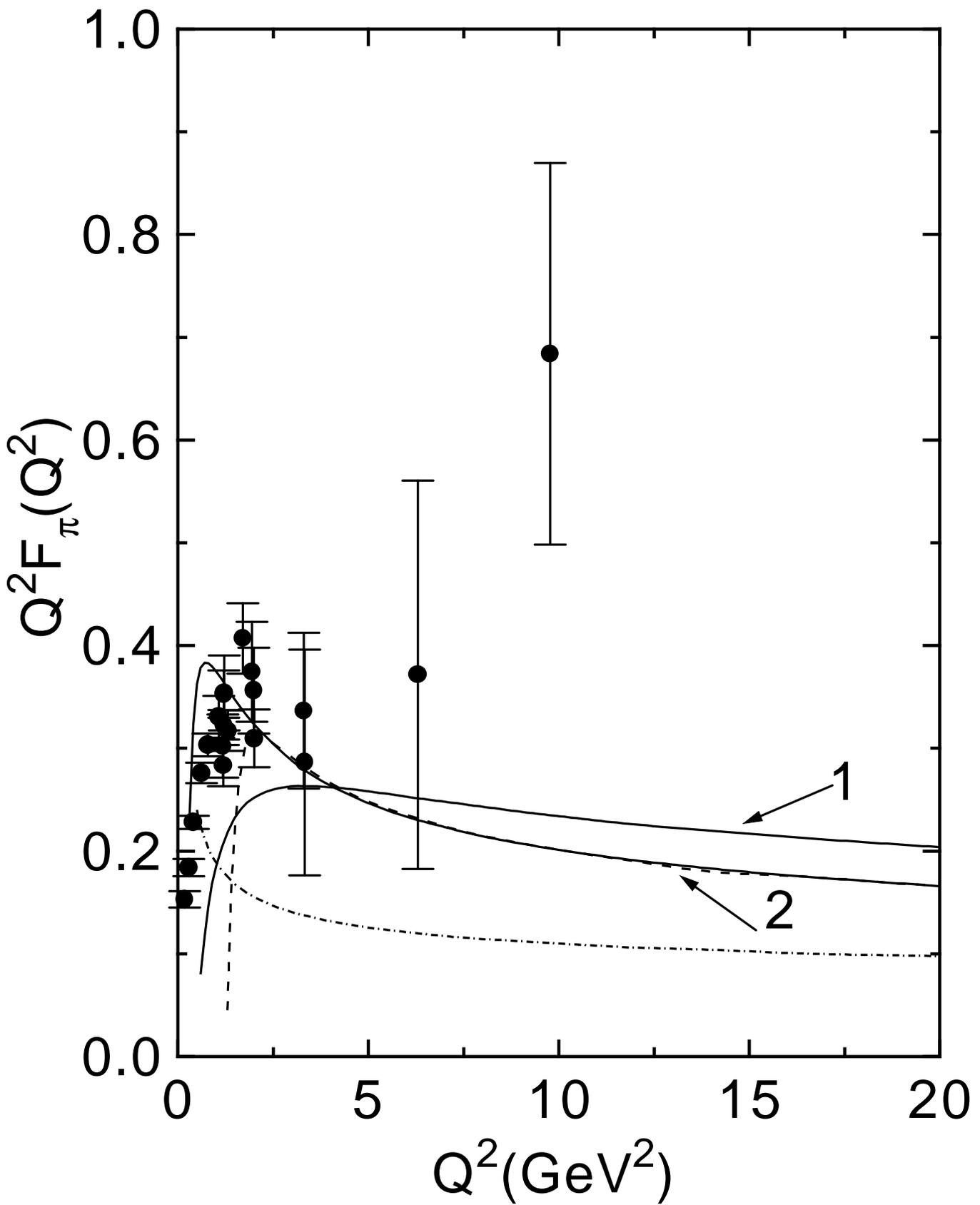
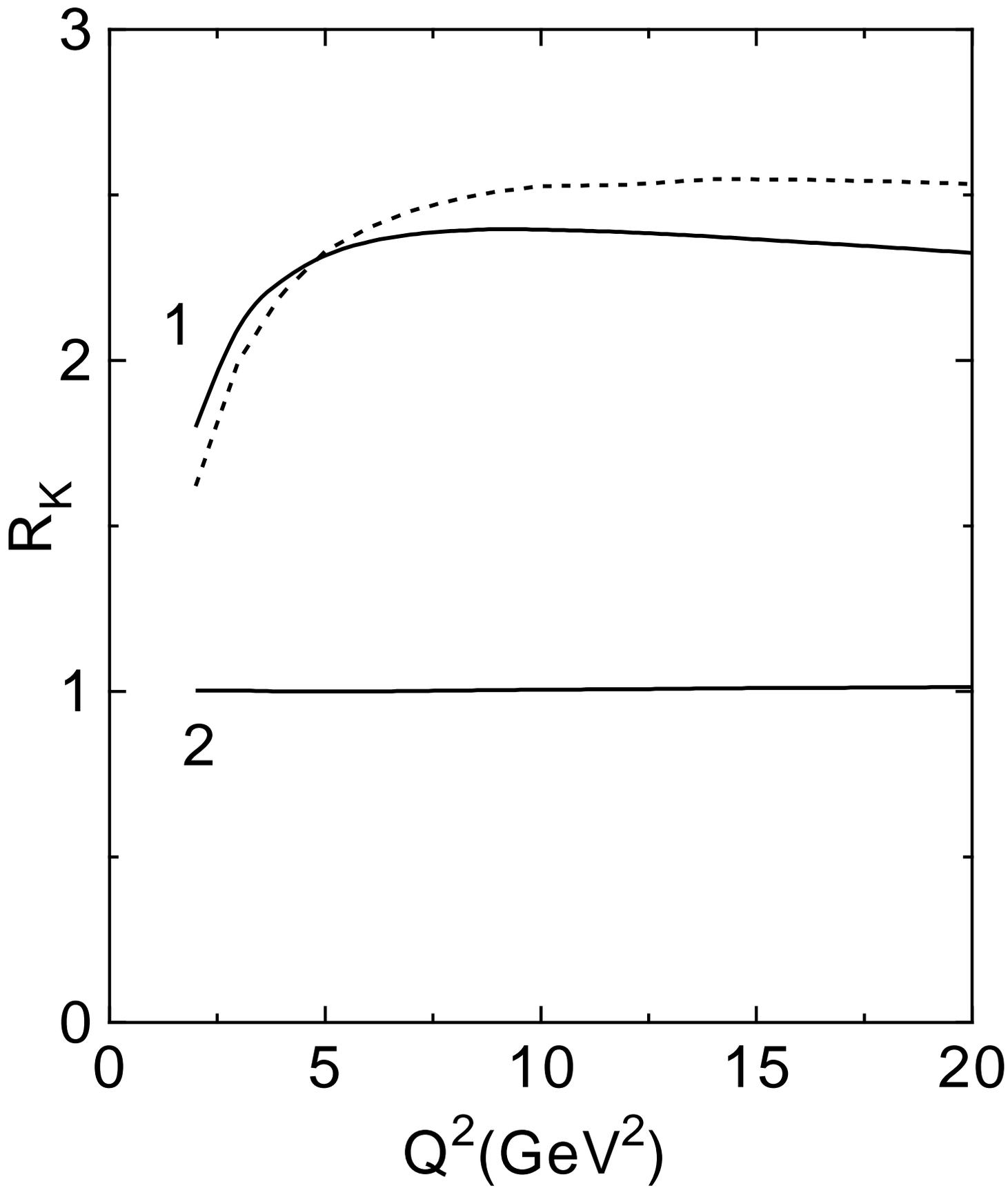
\end{document}